\documentstyle[spie1]{article}

\font\mrt=cmb10 at 16pt

\def\bendequ#1#2{\begin{equation}\label{#1}#2\end{equation}}
\def\parc#1#2{\frac{\partial #1}{\partial #2}}
\def\RE{{\rm Re}}
\def\V#1{{\bf #1}}

\def\dj{d\kern-0.8 ex\vrule 
height 1.32 ex depth -1.24 ex width 0.7ex\kern 0.15 ex}
\def\Dj{D\kern-1.65 ex\vrule 
height 0.95 ex depth -0.87 ex width 0.8ex\kern 0.78 ex}
\def\M {m}
\def\jen#1#2{\frac{{\rm d}\,#1}{{\rm d}\,#2}}
\def\dva#1#2{\frac{{\rm d}^2\,#1}{{\rm d}\,{#2}^2}}
\def\d={=}
\def\RE{\Re} \def\IM{\Im} \def\mm{n}
\def\half{{\footnotesize\frac{1}{2}}}
\def\quarter{{\footnotesize\frac{1}{4}}}
\def\dint{{\rm d\,}}
\def\q0{q_0}
\def\veferm{v_F}
\def\sign{{\rm sign}\,}
\def\textind#1{\indent\llap{\hbox to \parindent{#1\hfill}}\ignorespaces}

\begin{document}

\title{\mrt Quasiparticle Spectrum of Vortices in Cuprates}
\author{\large Irena Kne\v zevi\'c and Zoran Radovi\'c\\[8pt]
Department of Physics, University of Belgrade, P. O. Box 368,
 11001 Belgrade, Yugoslavia}
\date{}
\maketitle
\section*{ABSTRACT}

The quasiparticle excitation spectrum of an isolated vortex in 
a clean layered $d$-wave superconductor is calculated, using both the Bogoliubov--de Gennes and the quasiclassical Eilenberger equations. Analytical solutions are obtained within the model of step-variation of the gap function, adjusting the Caroli, de Gennes and Matricon approach for low-lying excitations in cuprates. A large peak in the density of states in the "pancake" vortex core is found, 
caused by the two-dimensional and strong coupling nature of high-temperature superconductivity. 

\vskip \baselineskip
\noindent
Keywords: vortex core, bound states, $d$-wave superconductivity

\section{INTRODUCTION}

Enhancing interest has been focused on quasiparticle spectra properties around vortices in both conventional and unconventional superconductors. Density of localized states near the Fermi surface in the vortex core in conventional type-II superconductors is equivalent to that of the normal metal area with radius of the order of the coherence length $\xi\sim 10\,{\rm nm}$.\cite{Dezen,Gygi,Bardin} High-temperature superconductors (HTS)
are layered, having a cylindrical Fermi surface, 
the superconductivity is of $d$-wave and strong coupling type, the coherence length being much smaller, $\xi\sim 1{\rm nm}$. This implies a crucial difference in the quasiparticle spectra in the vortex core between conventional superconductors and HTS. 

Numerous computations of the quasiparticle spectra around two-dimensional vortices have been performed recently. For $s$-wave pairing, Rainer {\it et al.} used the quasiclassical theory\cite{Rajner} and Hayashi {\it et al.} self-consistently solved the Bogoliubov--de Gennes equations in the quantum limit.\cite{Hayashi} For $d$-wave pairing, Schopohl and Maki\cite{Maki}  and Ichioka {\it et al.}\cite{Ichioka} analyzed the vortex structure within the quasiclassical theory, and Morita {\it et al.}\cite{Maki1} solved the Bogoliubov--de Gennes equations, neglecting the "quantum effects", i.e. non-commutativity between the quasiparticle momentum and coordinate, obtaining the four-fold symmetry of the vortex core. A fully self-consistent numerical approach, including "quantum effects", has been done by Morita {\it et al.}\cite{Maki2} and Franz and Te\v sanovi\'c.\cite{Franz} The common feature of layered  superconductors is the appearance of a large peak in the quasiparticle density of states (DOS) in the vortex core. This is in agreement with the scanning tunneling microscopy (STM) observations on ${\rm Nb Se_2}$, done by Hess {\it et al.},\cite{Hess} and on YBCO by Maggio-Aprile {\it et al.}.\cite{Magio} However, in BSCCO the gap structure was found by Renner {\it et al.},\cite{Renner} without any signature of the quasiparticle bound states in the vortex core. Possible explanation is either the existence of the pseudogap in the normal metal phase,\cite{Renner} or the low-lying states are not localized but extended along the gap node directions.\cite{Hess,Volovik}

Since the Andreev bound states can transport charge currents, unlike the bound states in a potential well, 
supercurrents can flow through the vortex, strongly influencing its dynamics.\cite{Rajner} 
In particular, this could be very important for understanding anomalous electric and thermal Hall conductivity
\cite{Krishana,Bozovic,Nagaoka} and the quantum magnetic de Haas--van Alphen oscillations.\cite{Otterlo} 

In this paper, both the Bogoliubov--de Gennes and the Eilenberger quasiclassical equations  are solved analytically, within the model of step-variation of the gap function, connecting the bound states in the vortex core and in the normal metal cylinder embedded in the same superconductor. Since the bound states spectrum is not quasicontinuous, for 2D and strong coupling superconductivity, this approach modifies the Caroli, de Gennes and Matricon results for low-lying states. Our results indicate that DOS has either one large maximum or a gap structure, depending on the value of $k_F\xi$.

\section{BOGOLIUBOV--DE GENNES EQUATIONS}

An efficient method for calculating local spectral properties is based on the Bogoliubov-de Gennes equations
\begin{eqnarray}\label{B-dG1}
\left\{\frac{1}{2\M}\left(-\imath\hbar\V\nabla-\frac{e}{c}\V A\right)^2
-E_F\right\}u+\Delta v&=&E u ,\nonumber\\
-\left\{\frac{1}{2\M}\left(-\imath\hbar\V\nabla+\frac{e}{c}\V A\right)^2
-E_F\right\}v+\Delta^* u&=&E v ,
\end{eqnarray}
where $u$ and $v$ are the Bogoliubov 
amplitudes and $\Delta$ is the gap function.\cite{Tinkham}  

For a normal-metal cylinder of radius $r_c$, embedded in 
a layered superconductor along the $z$-axis and in the absence of magnetic field, $\V A=0$, in cylindrical coordinates $(r,\varphi,z)$, $z\parallel c$, 
we assume the gap function in the form 
\bendequ{B-dG2}
{\Delta=\Delta(r,\theta)=\cases{0,& $r<r_c$\cr \Delta (\theta),& $r>r_c$},}
where 
\bendequ{dspar}
{\Delta(\theta)=\Delta_0\cos (2\theta)} 
for $d$-wave pairing. The self-consistency condition and non-commutativity between coordinate and momentum components are neglected, $\theta$ being the polar angle in $\V k$-space.\cite{Maki1} In this case, 
Eqs. (\ref{B-dG1}) can be rewritten in the form
\bendequ{B-dG14}
{\pmatrix{\hat L +E & -\Delta (r,\theta)\cr  \Delta (r,\theta) & \hat L -E }\pmatrix{f_+ \cr f_-}=0,}
where
\bendequ{B-dG12}
{\hat L= \frac{\hbar^2}{2\M}\left(\dva{}{r}+\frac{1}{r}\jen{}{r}-\frac{\mm^2}{r^2}+k_F^2\right)}
and
\bendequ{B-dG3}
{\pmatrix{u(\V r,\theta) \cr v(\V r,\theta)}=
\pmatrix{f_+(r,\theta)\cr f_-(r,\theta)}e^{\imath \mm\varphi},}
with $\mm=0,\pm 1,\pm 2,...$.

For $r<r_c$, Eq. (\ref{B-dG14}) has the following real-valued solution, finite at $r=0$ 
\bendequ{B-dG10}
{f_\pm=A_\pm J_\mm\left( q_{\pm} r \right).}
Here, $J_\mm$ is the Bessel function of the first kind, $A_\pm$ are   
arbitrary real numbers, and $q_{\pm}\d= 
\sqrt{k_F^2\pm 2\M E/\hbar^2}$. 

For $r>r_c$, the solution can be chosen in the form 
\bendequ{eigen}
{f_\pm =\sum_\lambda C_\pm^\lambda f_\lambda,}
where $\lambda$ and $f_\lambda$ are the eigenvalues and the corresponding eigenfunctions of $\hat L$,
\bendequ{eigen1}
{\hat L f_\lambda=\lambda f_\lambda.}

For bound states with energies 
below the superconducting gap, $|E| <|\Delta(\theta)|$
\bendequ{lambda}
{\lambda=\pm\imath\sqrt{\Delta^2 (\theta) -E^2},}
and $f_{\lambda}$ are 
the  solutions of the Bessel equation
\bendequ{B-dG17}
{\dva{f_{\lambda}}{r}+\frac{1}{r}\jen{f_{\lambda}}
{r}-\frac{\mm^2}{r^2}f_{\lambda}
+(k_F^2\mp\imath\q0^2)f_{\lambda}=0, }
with $\q0^2=(2\M/\hbar^2)\sqrt{\Delta^2 (\theta) -E^2}$. 
Introducing 
\bendequ{kdef}
{k\d= \sqrt{k_F^2+\imath\q0^2}\qquad (\RE k>0),} 
the real-valued solutions, which tend to zero as $r\to \infty$, are 
\bendequ{B-dG18}
{f_+=BH_\mm^{(1)}(kr)+B^*H_\mm^{(2)}(k^*r),}
\bendequ{B-dG19}
{f_-=-\frac{\imath\sqrt{\Delta^2 (\theta) -E^2}}{\Delta(\theta)}\left(BH_\mm^{(1)}(kr)-
B^*H_\mm^{(2)}(k^*r)\right)
+\frac{E}{\Delta(\theta)}\left(BH_\mm^{(1)}(kr)+
B^*H_\mm^{(2)}(k^*r)\right).}

Continuity of $f_+$, $f_-$ and their first derivatives at $r=r_c$ 
yields the system of algebraic equations for $A_\pm$, $\RE B$ and $\IM B$, 
which  has a nontrivial solution if
\bendequ{B-dG19a}
{\det {\pmatrix
{\ &\ &\ &\  \cr
J_\mm(+)&0 &-\RE H_\mm^{(1)}&\Im H_\mm^{(1)}\cr
\ &\ &\ &\  \cr
0& J_\mm(-)&-\frac{{\cal E}}{\Delta(\theta)}\IM H_\mm^{(1)}-
\frac{E}{\Delta(\theta)}\RE H_\mm^{(1)}&
-\frac{{\cal E}}{\Delta(\theta)}\RE H_\mm^{(1)}+
\frac{E}{\Delta(\theta)}\IM H_\mm^{(1)}\cr
\ &\ &\ &\  \cr
q_+{J_\mm}'(+)&0& 
-\RE (k{H_\mm^{(1)}}')&\IM (k{H_\mm^{(1)}}')\cr
\ &\ &\ &\  \cr
0&q_-{J_\mm}'(-)& 
-\frac{{\cal E}}{\Delta(\theta)}\IM (k{H_\mm^{(1)}}')-\frac{E}{\Delta(\theta)}
\RE (k{H_\mm^{(1)}}') &
-\frac{{\cal E}}{\Delta(\theta)}\RE (k{H_\mm^{(1)}}')+\frac{E}{\Delta(\theta)}\IM (k{H_\mm^{(1)}}')\cr 
\ &\ &\ &\  }}=0.}
Here ${\cal E}=\sqrt{\Delta^2 (\theta) -E^2}$, $J_\mm(\pm)=J_\mm(q_{\pm}r_c)$, $H_\mm^{(1)}=H_\mm^{(1)}(kr_c)$,  
$J'(x)=\dint J(x)/\dint x$, $H'(x)=\dint H(x)/\dint x$. Bound state energies are calculated numerically from Eq. 
(\ref{B-dG19a}).

For an isolated "pancake"  vortex in HTS in a low magnetic field, $H\sim H_{c1}\ll H_{c2}$, applied along the $z$-axis, we assume the same model for $\Delta (r,\theta)$,  Eq. (\ref{B-dG2}), taking $\V A\approx 0$ and the gauge
\bendequ{E8}
{\Delta=\Delta(r,\theta)e\sp{-\imath\varphi}.}
due to the flux quantization. The core radius $r_c$ is now $r_c(\theta)$, of the order of the BCS coherence length $\xi_0(\theta)=
\hbar v_F/\pi|\Delta(\theta)|$. Taking
\bendequ{B-dG21}
{\pmatrix{u(\V r,\theta) \cr v(\V r,\theta)}=
\pmatrix{f_+(r,\theta)e^{\imath(\nu-1/2)\varphi}\cr 
f_-(r,\theta)e^{\imath(\nu+1/2)\varphi}},}
where $2\nu$ is an odd integer, Eq. (\ref{B-dG14}) approximately describes the vortex core, if $\hat L\to \hat L '$ and $E\to E'$, where
\bendequ{B-dG23}
{\hat L '\d= \frac{\hbar^2}{2\M}\left(\dva{}{r}+\frac{1}{r}\jen{}{r}-\frac{\nu^2+\quarter}{r^2}+k_F^2\right) }
and
\bendequ{B-dG23a}
{E'\d= E+\frac{\hbar^2|\nu|}{2\M r_c^2(\theta)}\sign E}
describe the influence of the screening currents flow around the vortex on the low-lying bound states. In this way the solution for the vortex core is related to the solution for normal metal cylinders with $\theta$-dependent $r_c$. This approximation is close to the approach in Ref. [1], adjusted for HTS, where only one or two lowest bound states appear. 

Density of bound states
\bendequ{dos}
{N_b(E)=\frac{1}{2\pi}\int_0^{2\pi}\dint\theta\sum_i
\delta\left(E-E_i (\theta)\right),}
where $i$ is $n$ for a cylinder and $\nu$ for a vortex, is calculated approximately for $E>0$, taking  for a cylinder
\bendequ{approx}
{{E_n}(\theta)\approx {E_n}(0)|\cos(2\theta)|,}
and for a vortex
\bendequ{approx1}
{{E_\nu}(\theta)\approx {E_\nu}'(0)|\cos(2\theta)|-\frac{\hbar^2|\nu|}{2\M r_c^2(\theta)}.}
Therefore, for a cylinder
\bendequ{cyl}
{N_{b}(E>0)=\frac{2}{\pi}\sum_n\frac{\Theta(E_n(0)-E)}{\sqrt{E_n(0)^2-E^2}}}
and for a vortex
\begin{eqnarray}\label{vort1}
&&N_{b}(E>0)=\frac{2}{\pi}\sum_\nu 
\frac{\Theta\left(\alpha^2-4E\beta\right)}{\sqrt{\alpha^2-4E\beta}}
\times\nonumber\\ 
&&\times\left[
\frac{\Theta\left(\frac{\alpha+\sqrt{\alpha^2-4E\beta}}{2\beta}\right)
\Theta\left(1-\frac{\alpha+\sqrt{\alpha^2-4E\beta}}{2\beta}\right)}
{\sqrt{1-\left(\frac{\alpha+\sqrt{\alpha^2-4E\beta}}{2\beta}\right)^2}}
+\frac{\Theta\left(\frac{\alpha-\sqrt{\alpha^2-4E\beta}}{2\beta}\right)\Theta\left(1-\frac{\alpha-\sqrt{\alpha^2-4E\beta}}{2\beta}\right)}
{\sqrt{1-\left(\frac{\alpha-\sqrt{\alpha^2-4E\beta}}{2\beta}\right)^2}}\right],\end{eqnarray}
where $\alpha=E_\nu '(0)$, and $\beta=|\nu |\hbar^2/2\M\xi(0)^2$. Low-lying bound states are well described by taking the constant energy shift in Eq. (\ref{approx1}), whereas for larger $\nu$ the shift becomes zero, $E'(\theta)\approx E(\theta)$, as shown by perturbative calculations.\cite{mi1}

\section{EILENBERGER EQUATIONS}

In this section, we consider the same model in the quasiclassical approach. We start with the Eilenberger equations in the following form\cite{Rajner,Maki}
\bendequ{E4}
{\left[ 2\hbar\omega\sb n +\hbar {\V v}_F\cdot\left(\V \nabla -\imath \frac
{2e}{\hbar c}\V A\right)\right] f=
2\Delta g,\quad \left[ 2\hbar\omega\sb n -\hbar {\V v}_F\cdot\left(\V \nabla +\imath\frac
{2e}{\hbar c}\V A\right)\right] f\sp{\dagger}=
2\Delta\sp * g, }
\bendequ{E6}
{\hbar{\V v}_F\cdot\V \nabla g=\Delta\sp *f-f\sp{\dagger}\Delta .}
Here $g=g_{\downarrow\downarrow}(\V r,{\V v}_F,\omega\sb n)$ and $f=f_{\downarrow\uparrow}(\V r,{\V v}_F,\omega\sb n)$ represent the normal and the anomalous Green function  respectively, $\Delta =\Delta (\V r,{\V v}_F)$ is the gap function, $\omega\sb n=\pi k\sb B T(2n+1)$, $n=0,\pm 1,\pm 2...$, 
are the Matsubara frequencies, and ${\V v}_F$ is the  Fermi velocity vector. The function  $f\sp{\dagger}$ is defined by $f_{\uparrow\downarrow}\sp \dagger(\V r,{\V v}_F,\omega\sb n)=f_{\downarrow\uparrow}\sp *(\V r,-{\V v}_F,\omega\sb n)$. $\Delta$ and $f$ are connected by the self-consistency equation. 

For a homogeneous and isotropic superconductor, solutions of 
the Eilenberger equations are
\bendequ{E3a}
{\langle f\rangle=\frac{\Delta}{\varepsilon_n},
\quad \langle f\sp{\dagger}\rangle=
\frac{\Delta\sp*}{\varepsilon_n},
\quad \langle g\rangle =\frac{\hbar \omega\sb n}{\varepsilon_n},}
where 
\bendequ{E3b}
{\varepsilon_n\sp 2=\vert \Delta\vert\sp 2+(\hbar\omega\sb n)\sp 2.}

For a pancake vortex in ($r,\varphi $) plane, using the model gap function, Eqs. (\ref{B-dG2}) and (\ref{E8}), 
and denoting the coordinate along ${\V v}_F$ by $s$, and along  
$\hat z\times {\V v}_F$ by $p$ (Fig.1.), 
in the gauge with real-valued gap function, Eqs. (\ref{E4}) and (\ref{E6}) can be rewritten 
in the form\cite{mi2}
\bendequ{E13}
{\left[ 2\hbar\omega\sb n +\hbar \veferm \left(\parc{}{s} +\imath\frac{p}{r\sp 2}\right)\right] f=2\Delta (r,\theta)g,}
\bendequ{E13a}
{\left[ 2\hbar\omega\sb n -\hbar  \veferm \left
(\parc{}{s} -\imath\frac{p}{r\sp 2} \right)\right] f\sp{\dagger}=
2\Delta(r,\theta) g,}
\bendequ{E15}
{\hbar \veferm \parc{}{s}g=\Delta(r,\theta)(f-f\sp{\dagger}),}
where $r^2 =p^2 +s^2$.

\nopagebreak
\vskip 4cm
\noindent
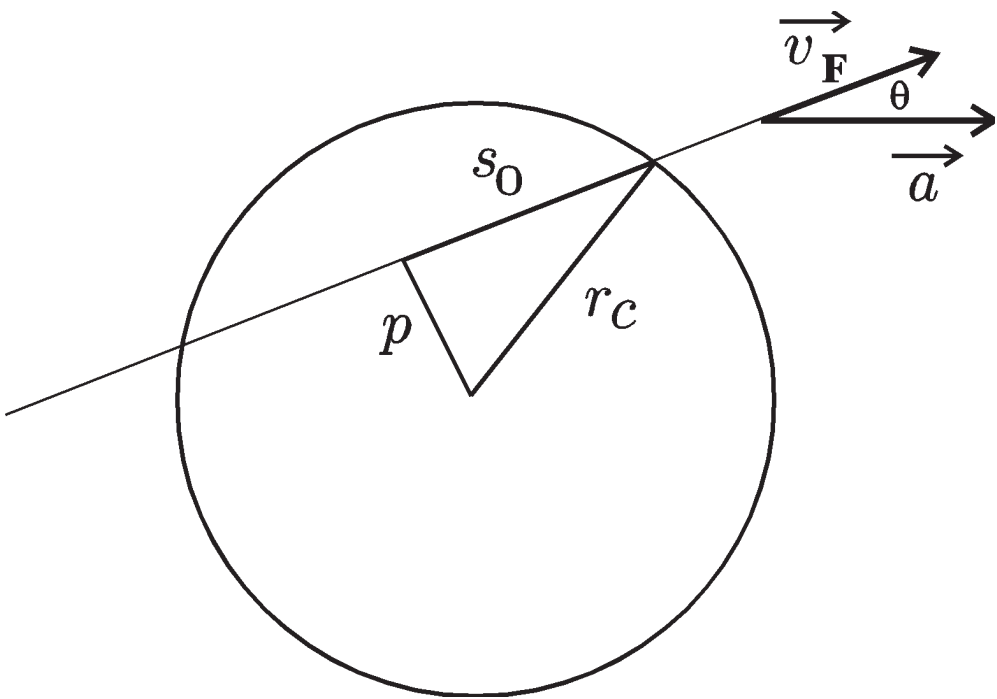
{{ Fig.1.} Trajectory passing at distance $p$ from the vortex center.\label{vort}}
 \vskip \baselineskip
Normal metal cylinder in zero magnetic field is described by Eqs. (\ref{E13})-(\ref{E15}) without $\imath p/r^2$ term. 
For $r< r_c$, 
\bendequ{E18}
{f=Fe\sp{-\kappa\sb 0 s},\quad  g=G,}
where $\kappa\sb 0=2\omega_n/\veferm$. For $r> r_c$ and  $\kappa=2\varepsilon_n/\hbar \veferm $
\bendequ{E19}
{f=\langle f\rangle +\Phi\sb 1 e\sp{-\kappa s},
\quad g=\langle g\rangle +\Gamma\sb 1 e\sp{-\kappa s},\quad\mbox{for $s>0$},}
\bendequ{E20a}
{f=\langle f\rangle +\Phi\sb 2 e\sp{\kappa s}
,\quad g=\langle g\rangle +\Gamma\sb 2 e\sp{\kappa s},\quad\mbox{for $s<0$.}}
Eqs. (\ref{E13})-(\ref{E15}) imply
\bendequ{E21}
{\frac{\Phi\sb 1}{\Gamma\sb 1}=\frac{\Delta (\theta)}{\hbar \omega_n -\varepsilon_n},\quad  
\frac{\Phi\sb 2}{\Gamma\sb 2}=\frac{\Delta (\theta)}{\hbar \omega_n +\varepsilon_n}.}
From the continuity condition for $f$ and $g$ at 
$\pm s\sb 0=\pm \sqrt{r_c^2-p^2}$, for $r<r_c$ the normal Green function is 
\bendequ{E25}
{G=\frac{\hbar\omega_n\cosh (\kappa_0 s_0)+\varepsilon_n \sinh (\kappa_0 s_0)}
{\hbar\omega_n\sinh (\kappa_0 s_0)+\varepsilon_n \cosh (\kappa_0 s_0)}.}

For a vortex, approximating $p/r^2$ by $p/r_c^2(\theta)$, 
the solution of Eq. (\ref{E13})-(\ref{E15}) 
can be obtained from Eq. (\ref{E25}), by changing 
$\omega_n\to{\omega_n}'$, 
\bendequ{Gp}
{G\approx \frac{\hbar\omega_n '\cosh (\kappa_0 's_0)+\varepsilon_n '\sinh (\kappa_0 's_0)}
{\hbar\omega_n '\sinh (\kappa_0 's_0)+\varepsilon_n '\cosh (\kappa_0 's_0)},}
where 
\bendequ{E25a}
{{\omega_n}'=\omega_n +\frac{\imath p\veferm }{2r_c^2(\theta)},}
and $\kappa_0 '=2\omega_n '/\veferm $. 

Performing an analytical continuation of $G$ by $\hbar\omega_n\to -\imath E +0$, $E$ being the quasiparticle energy with respect to the Fermi level, the retarded propagator $g^R(E, p, \theta )$ is obtained.  

For a normal-metal cylinder, the bound states spectrum  is given by
\bendequ{spectrum}
{E=\left|\Delta(\theta)\cos\left(\frac{2Es_0}{\hbar\veferm}\right)\right|{\rm sign}\,\left[\tan\left(\frac{2Es_0}{\hbar\veferm}\right)\right].}

For a vortex and low-lying states,\cite{mi1} Eq. (\ref{spectrum}) holds for $E\to E'$, 
\bendequ{spectvort}
{E'=E+\frac{\hbar |p|\veferm}{2r_c^2(\theta)}\sign E.}

In terms of  reduced variables $E/\Delta_0\to E$,
$\sqrt{\Delta^2(\theta)-E^2}/\Delta_0 \to \varepsilon$, $\sqrt{E^2 -\Delta^2(\theta) }/\Delta_0 \to e $,  $p/r_c\to p$, $2s_0/\pi \xi_0 \to s_0$, the angle-resolved partial DOS (PDOS) is obtained from $N(E, p, \theta )=\RE g^R(E, p, \theta )$. 

For the normal-metal cylinder
\begin{eqnarray}\label{E30}
N(E , p, \theta )/N(0)&=&\Theta \left(e ^2\right)
\frac{|E|e }{e ^2\cos^2 (Es_0)
+E^2\sin^2 (Es_0)}+\nonumber\\
&+&\Theta \left(\varepsilon^2\right)\frac{\pi |\Delta(\theta)|}{\Delta_0}\delta \left(E\sin (Es_0)-
\varepsilon\cos (E s_0)\right),\end{eqnarray}
where $\delta$ is the Dirac function, $\Theta$ is the step-function, and $N(0)=m/2\pi\hbar^2$ 
is the normal metal DOS for one spin orientation. Averaging PDOS over the cylindrical Fermi surface gives
\begin{eqnarray}\label{E39}
N(E,p )/N(0)&=&\frac{1}{2\pi}\int_0^{2\pi} N(E, p, \theta )/N(0)\, d\theta =\nonumber\\
&=&\frac{2}{\pi}\int_{\sqrt{\max \{ 0, E^2 -1\}}}^{|E|}\, 
\frac{|E|e ^2\, de }{\sqrt{E^2-e ^2}
\sqrt{1-E^2+e ^2}
\left(e ^2\cos^2(E s_0)+
E^2\sin^2(E s_0)\right)}+\nonumber\\
&+&\frac{\left(E \tan (E s_0)+|E \tan (E s_0)|\right)}
{\sqrt{\cos^2 (E s_0)-E^2}}
\Theta \left(\cos^2 (E s_0)-E^2\right). \end{eqnarray}
Finally, after spatial averaging over the cylinder area $\pi r_c^2$, DOS is 
\bendequ{E40}
{N(E)=\frac{4}{\pi}\int_0^1 N(E,p)\sqrt{1-p^2} \, dp.}

For the vortex, in Eq. (\ref{E30}) $E\to E+E_0\,{\rm sign}\, E$, 
$E_0={\hbar |p_\nu|\veferm }/{2\Delta_0r_c^{2}(\theta)}$, $p_\nu =\nu/k_F$, and Eqs (\ref{E39}) and (\ref{E40}) become more complicated. For small-radius vortices in HTS, $k_F\xi_0(0)\sim 1$, where only one trajectory through the vortex core is allowed, with $p=p_{\half}=1/2k_F$, the quasiclassical DOS, with one large peak, is qualitatively the same as for the normal-metal cylinder.

\section{RESULTS AND DISCUSSION}

Our results are illustrated in Figs. 2-6. for $\Delta_0/E_F=0.4$, corresponding to $k_F\xi_0(0)=2E_F/\pi\Delta_0=1.5$, which is a reasonable value for cuprates, providing one distinguishable peak in DOS in the vortex core. 

For a normal-metal cylinder of radius $r_c=5\xi_0(0)$ embedded in a layered $d$-wave superconductor, seven distinguishable localized states are found (Fig. 2). This number obviously decreases with approaching the gap function nodes in the ${\bf k}$-space. For a cylinder of radius $r_c=\xi_0(0)$, the minimal vortex core radius, only two bound states are distinguishable, as shown in Fig. 3. Note that the lowest level of the quasicontinuum bound states calculated from quasiclassical theory (dashed curve in Fig. 3) is higher than in the quantum approach. However, for a $d$-wave pancake vortex, only one localized state can be seen (Fig. 4.) in the core of radius $r_c(\theta)=\xi_0(\theta)$. Note that only this state ($\nu =1/2$) is shifted, according to the arguments from the more rigorous perturbation theory approach.\cite{mi1} 

Density of bound states, calculated within the quantum model, for the normal-metal cylinder and for the vortex core, is shown in Fig. 5. Only one peak in DOS appears (except for the usual discontinuity at the gap edge) for the chosen value $k_F\xi_0(0)=1.5$. In the quasiclassical approach the total DOS, including the unbound states, is calculated for the normal-metal cylinder for the same parameters and shown in Fig. 6. Qualitatively, this is the same as the dashed curve in Fig. 5, despite the fact that the quantization of bound states is not taken into account. Note that in the quasiclassical approach the peak energy is systematically shifted towards higher energies. 

In conclusion, 
the approximate analytical method is developed in the same manner as the Caroli, de Gennes and Matricon approach, adjusted for the case of small-radius pancake vortices, where the quasiparticle spectrum is not quasicontinuous. Two-dimensional and strong-coupling nature of superconductivity in cuprates implies that the density of localized quasiparticle states in the vortex core is rather different from the normal metal DOS. 
Comparing our results with the experimental data,\cite{Magio,Renner} it seems that only one large peak in DOS is distinguishable for $0.7<k_F\xi_0(0)<1.5$, as in YBCO, and none (gap structure) for $k_F\xi_0(0)<0.7$, as in BSCCO, in consistency with a larger gap for the latter.\cite{Harris} This can be the simplest explanation of the experimental observations. 

\clearpage
\hfill
\vskip -1\baselineskip
\vskip 5.8cm

\noindent
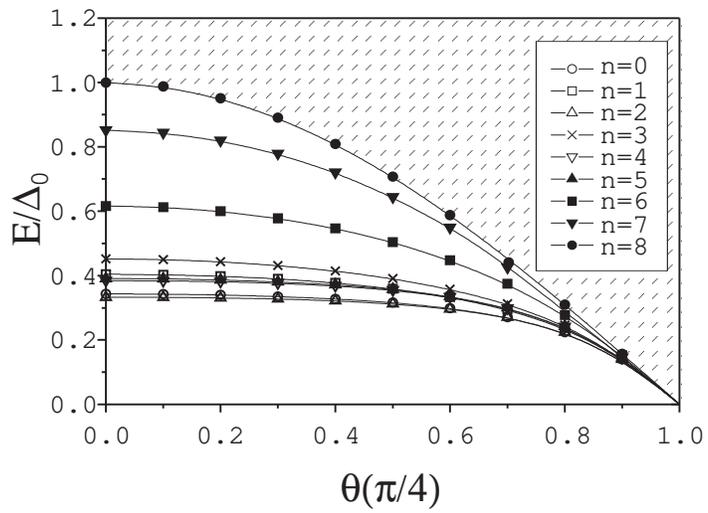
{Fig. 2. Quasiparticle energy $E/\Delta_0$ vs the momentum orientation angle in a normal-metal cylinder, 
calculated from Eq. (\ref{B-dG19a}), for $k_F\xi_0(0)=1.5$ and $r_c=5\xi_0(0)$. Shaded region: unbound states.}

\vskip \baselineskip
\vskip 5.58cm

\noindent
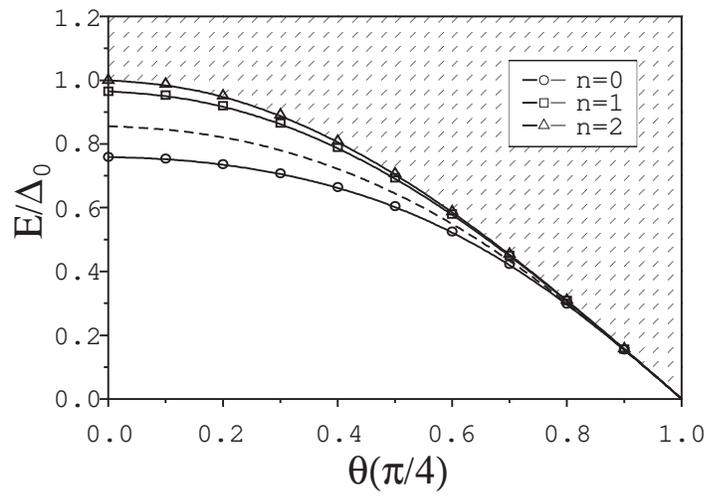
{Fig 3. Quasiparticle energy $E/\Delta_0$ vs the momentum orientation angle in a normal metal cylinder, 
calculated from Eq. (\ref{B-dG19a}), for $k_F\xi_0(0)=1.5$ and $r_c=\xi_0(0)$. Dashed curve: the lowest bound state 
($p=0$), calculated from Eq. (\ref{spectrum}) within the quasiclassical theory. Shaded region: unbound states.}

\vskip \baselineskip
\vskip 5.77cm

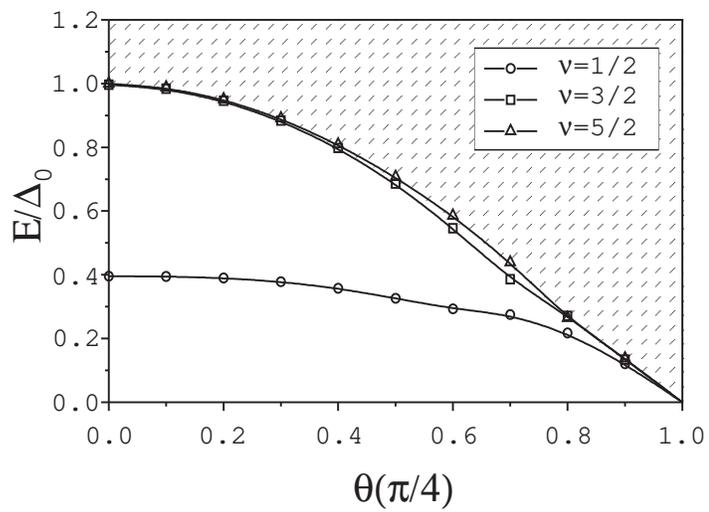
\noindent{Fig. 4. Quasiparticle energy $E/\Delta_0$  vs the momentum orientation angle in the vortex core 
of radius $r_c=\xi_0(\theta)$, calculated from Eq. (\ref{B-dG19a}), for $k_F\xi_0(0)=1.5$. Shaded region: unbound states.}

\clearpage
\hfill
\vskip -1\baselineskip
\vskip 5.43cm

\noindent
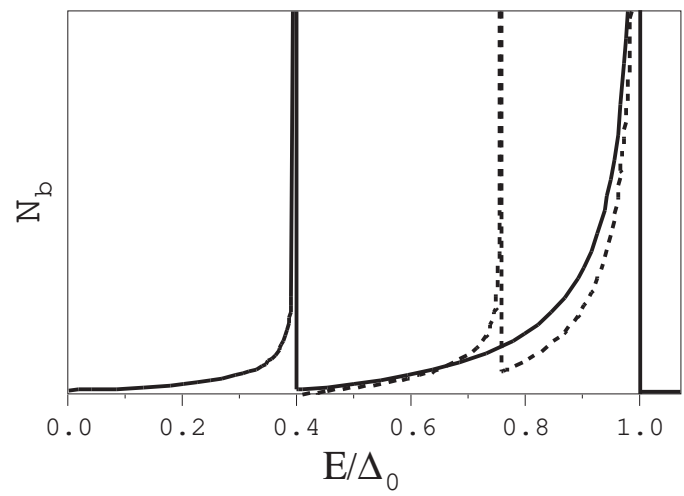
{Fig. 5. Density of bound states in the vortex core $r_c=\xi_0(\theta)$ 
(solid curve) and in the normal-metal cylinder $r_c=\xi_0(0)$ (dashed curve), 
calculated from Eqs. (\ref{cyl}) and (\ref{vort1}), for $k_F\xi_0(0)=1.5$.}

\vskip 4.99cm
\nopagebreak

\noindent
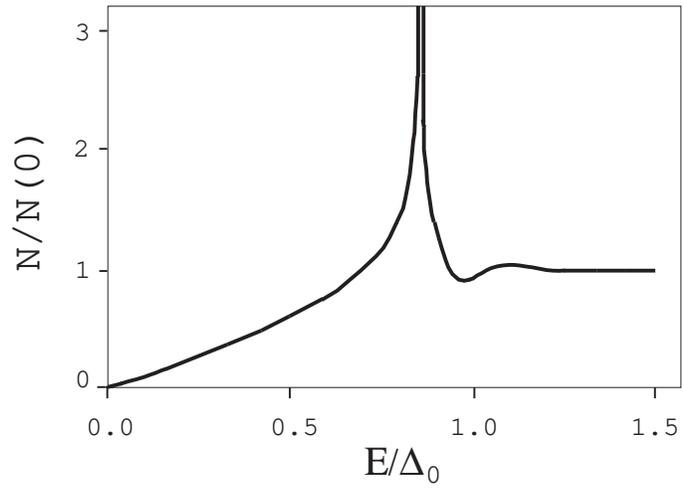
{Fig. 6. Quasiparticle DOS in the normal metal cylinder, calculated from Eq. (\ref{E40}), 
for $r_c=\xi_0(0)$  and $k_F\xi_0(0)=1.5$, within the quasiclassical theory.}

\end{document}